# Negative Friction and Reversible Energy Exchange Between Orderly Motion and Phonons in Carbon Nanotube Oscillators

Yang Zhao[1], Chi-Chiu Ma[1], GuanHua Chen[1,*], ZhiPing Xu[2], QuanShui Zheng[2,†], and Qing Jiang[3]

[1]*Department of Chemistry, University of Hong Kong, Hong Kong, P.R. China,*
[2]*Department of Engineering Mechanics, Tsinghua University, Beijing 100084, P.R. China,*
and [3]*Department of Mechanical Engineering, University of California, Riverside, U.S.A.*
(Dated: Septemper 4, 2004)

Energy exchanges between orderly intertube axial motion and vibrational modes are studied for isolated systems of two coaxial carbon nanotubes at temperatures ranging from 300 K to 500 K. It is found that the excess intertube van der Waals energy, depleted from the intertube axial motion, is primarily stored in low-frequency mechanical modes of the oscillator for an extended period of time, and furthermore, such an energy exchange may be reversible. A new nanoscale tribological phenomenon of negative friction is reported for the first time in concentric cylindrical carbon nanotube oscillators.

Concentric cylindrical carbon nanotubes have been recently proposed [1], and intensely studied [2–6], as promising candidates for nanoscale molecular bearings, springs, and oscillators. Performance and load-bearing properties of fundamental components of nanomachines have yet to be understood despite their unlimited prospects of applications. In addition, nanomachinery has also been suggested to serve as a test bed for ergodicity and equipartition on complex energy surfaces[3]. In fact, coupled oscillator systems have undergone a large number of investigations as model systems for studying energy exchanges amongst various degrees of freedom, ergodicity on energy surfaces, and equipartition as systems relax. Fermi, Pasta, and Ulam reported the first numerical experiment on a chain of 32 coupled oscillators with quartic anharmonicity in 1954, which is often referred to as the Fermi-Pasta-Ulam (FPU) model[7]. For a particular initial energy distribution it is observed that the oscillators do not relax to an equipartition state for an extended period of time, instead, they display a persistent recurrence to the initial condition, seemingly contrary to the equipartition hypothesis of statistical mechanics. To our knowledge, similar types of recurrence representing a tidal, reversible energy exchange amongst two or more dominant modes, however, have not been reported for systems of much higher dimensionalities.

We report in this Letter an interesting, counterintuitive energy exchange phenomenon, which takes place between the orderly intertube axial oscillatory motion and mechanical modes in double-walled nanotube (DWNT) oscillators, and which is found to be *reversible*. What makes this particular form of energy exchange appealing is that the intertube axial oscillation gradually dissipates the excess intertube van der Waals energy, created by the initial inner tube extrusion, into various vibrational modes in the DWNT, and at the crossroad of the energy dissipation process lies this reversible energy exchange between the excess intertube van der Waals energy and the DWNT low-frequency mechanical modes. The reversibility of this particular form of energy exchange is especially interesting as the intertube axial oscillatory motion is considered to be a form of orderly motion whose revival may point to local loss of entropy. Also of interest is that the time scale of the axial motion is comparable to that of low-frequency mechanical modes in the DWNT, which renders energy transfer between the two more obtainable but nonetheless intriguing.

Legoas *et al.* run simulations with a canonical ensemble for a variety of temperatures up to 400 K, and Rivera *et al.*, for a temperature range from 275 to 450 K [8, 9]. Guo *et al.* thermally equilibrate DWNT oscillators with a bath to reach an initial temperature $T_i$, then switch to a microcanonical ensemble for simulations [10], and a similar approach is used by Servantie and Gaspard fixing $T_i$ at 300 K [11]. Similar to the nano-oscillator setup in our previous work [3], the system is chosen to be a DWNT. The DWNT construct is one of the most elementary realizations of a nanoscale oscillator. In this Letter, we concentrate on one particular configuration: the outer and inner tubes both chosen to be of the zigzag type, specifically, the open-ended outer tube is (14,0) with a length of 70 Å, and the capped inner tube is (5,0) with a length of 55 Å.

Here we take a similar approach as that of Guo *et al.* [10] In simulation performed here, the geometry of the oscillator is optimized first, and then the DWNT is heated up to 300K $\sim$ 600K for 20ps, and thermally equilibrated for 200ps, before commencing simulations in a microcanonical ensemble. After the thermal equilibration, the inner tube is displaced by 35 Å such that the initial extrusion length of the inner tube s = 27.5 Å. Simulation of the DWNT oscillation is then carried out using the CHARMM force field, and various energies are calculated as a function of time. A time step of 1 fs was used for all simulations.

In Fig. 1, the intertube axial oscillation amplitude as represented by center-of-mass distance between two nanotubes is displayed as a function of time for up to 1000 ps. As shown in the upper (lower) panel, the inner tube was released from an initial extrusion length of s=27.5 Å to commence intertube axial oscillations after the DWNT oscillator has been heated up to 300K (500K), and due to pre-simulation heating and dynamic intertube roughness at atomic scales, intertube axial oscillation is rapidly damped in the first 500 (400) ps. As displayed in the upper (lower) panel, from t = 500 (400) to 800 (750) ps, the intertube axial oscillation is virtually diminished, and the excess intertube van der Waals energy is transfered to other forms of energies which are to be identified. To our surprise, from t=800 (750) ps the DWNT starts to oscillate again, albeit with reduced amplitudes. We call the period between t=500 (400) ps and 800 (750) ps the hibernation period, and the time after t=800 (750) ps, the awakening period. Detailed examinations of the hibernated DWNT oscillator reveal the low-energy vibrational modes are peaked around 0.5 THz as shall become clearer.

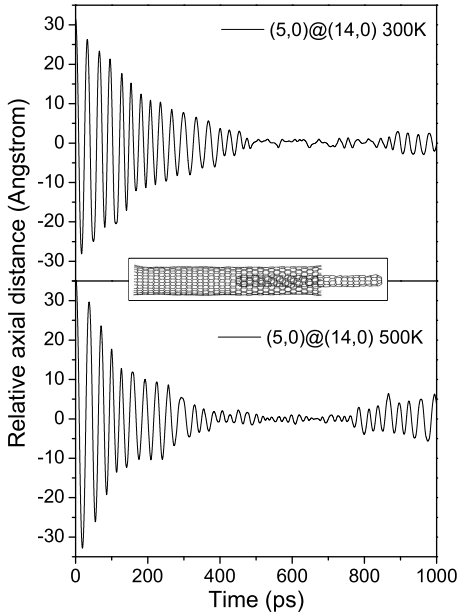

FIG. 1: The intertube axial oscillation amplitude as represented by center-of-mass distance between two nanotubes as a function of time. Upper panel: s=27.5 Å, $T_i = 300K$. Lower panel: s=27.5 Å, $T_i = 500K$. Inset shows the DWNT oscillator.

Fig. 2 shows a frequency-domain analysis of the radial movements of a carbon atom in the center portion of the outer tube before, during, and after the hibernation period of the intertube axial oscillation. The radial movements during the hibernation period are attributed to low-frequency vibrations of the DWNT with both tubes bending or waving. Such vibrations with resonance frequencies in the tera-hertz range have been previously reported [12]. As a result, the center portion experiences larger radial motion than the end portion. For atoms in the center portion of the outer nanotube, radial oscillations with an approximate period of 2 ps exist for all three time periods, but the radial oscillation amplitude is the largest during the hibernation period of intertube axial oscillation implying a significant energy transfer into DWNT mechanical modes during that time period. In the insets of Fig. 2, corresponding time-domain pictures are shown for the radial movements of the carbon atom located in the center portion of the outer nanotube before, during, and after the hibernation period.

In addition to the aforementioned bending-waving motion, the DWNT oscillator also experiences an intertube angular motion prior to and during the hibernation period. The period of this rotational motion is about 20 ps, and due to energy exchanges, the rotational speed also fluctuates. In the top panel of Fig. 3, the velocity of intertube axial motion for the case of s=27.5 Å and $T_i = 300$ K is shown for the first 1 ns. In comparison, the intertube relative angular velocity is displayed in the second panel in Fig. 3. The relative angular velocity peaks when the hibernation period commences at t = 500ps. Further proof is provided in the third and fourth panels in which the kinetic energies of the intertube axial motion and relative rotational motion are displayed, respectively. Therefore, in addition to bending-waving motion, the relative angular motion is also present in the hibernation period, and on a declining slope. At t=1 ns, the intertube rotation comes to a near stop with the intertube axial motion and other low-frequency mechanical modes being the energetic beneficiaries. For longer times beyond t=1 ns, our simulation shows that the hibernation period of the intertube axial motion can reappear as energies are again transferred back into the bending-waving motion and the rotational modes.

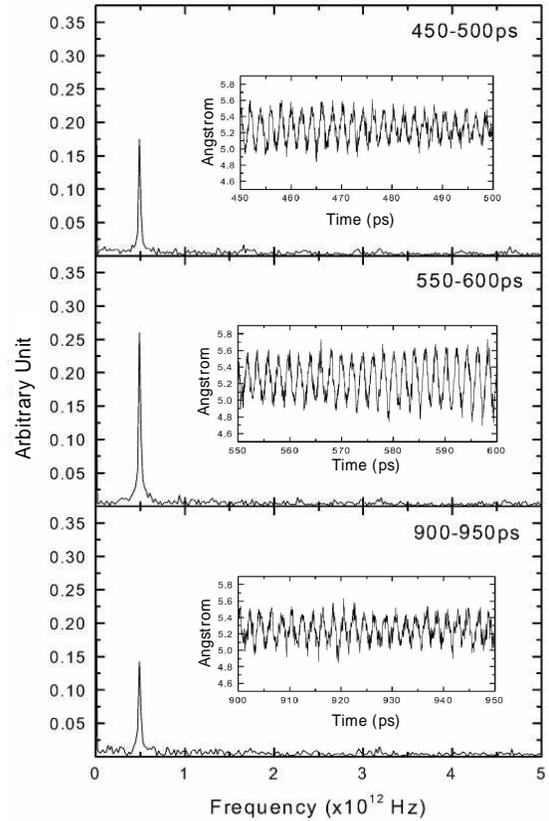

FIG. 2: Frequency-domain analysis of the radial movements of a carbon atom located in the center portion of the outer nanotube before (upper panel), during (middle panel), and after (lower panel) the hibernation period of the intertube axial oscillation. Initial extrusion s=27.5 Å, and $T_i = 300$ K. Insets: corresponding time-domain pictures for the radial movements of the same carbon atom.

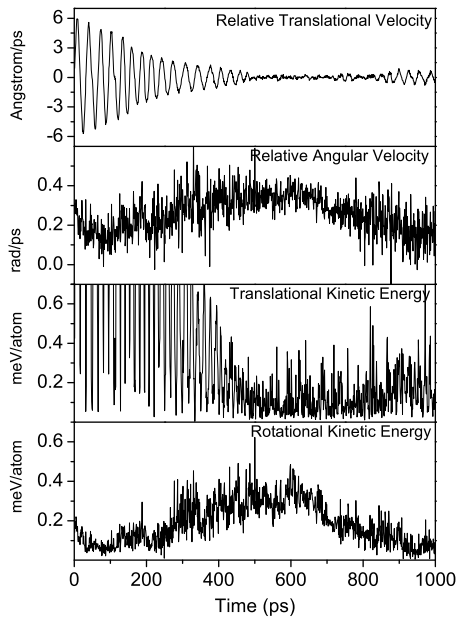

FIG. 3: The intertube axial velocity (first panel) and the corresponding kinetic energy (second panel), and the relative angular velocity (third panel) and the corresponding kinetic energy (fourth panel), are plotted for s= 27.5 Å, $T_i = 300K$. The first 1000 ps of the simulation is shown.

Quite different definitions of the intertube frictional force in the axial direction have been proposed for the DWNT oscillators [3, 11, 13]. For example, frictional forces can be derived from the time-dependent changes of the intertube center-of-mass velocity. Energetic considerations of the frictional force can also be contemplated in place of kinetic ones. Following our previous work [3], here we estimate the frictional force per carbon atom from energy decay rates of the intertube axial oscillation. The outer tube has a larger mass than the inner tube, and therefore, it has a smaller displacement by the conservation of the system momentum. Therefore the speed of the inner tube can be a rough estimate of the intertube speed. For s=27.5 Å and $T_i$=300K, the average velocity of the inner tube from t= 0 ps to 450 ps is 1.9 Å/ps, and the frictional force estimated for the same time period is about $1.8 \times 10^{-14}$ N per atom. However, starting from the hibernation period, due to the transfer of energy stored in the vibrational-rotational modes back to the intertube axial oscillation, the frictional force thus estimated between 600 and 1000 ps is $-1.7 \times 10^{-17}$ N per atom. The fact that the friction is negative is a direct result of non-ergodicity in the DWNT oscillator system. Similar to the persistent recurrence to the initial condition in the seminal work of Fermi, Pasta, and Ulam, reversible energy transfers between the intertube axial motion and low-frequency mechanical modes, such as the intratube bending-waving and rotational modes, reveal difficulties for a classical system of a miniature size to relax expeditiously to an equipartition state.

Simulations on an armchair DWNT oscillator with a (7,7) inner tube and a (12,12) outer tube have also been carried out. Incomplete hibernation of the intertube axial motion has been found in which the oscillatory translational motion is significantly reduced while its energies are transferred into various low-energy mechanical modes in the armchair DWNT oscillator. But unlike in the zigzag DWNT oscillator, energies have largely gone into relative rotational modes from the partially hibernated intertube axial motion in the armchair DWNT oscillator. As the oscillator recovers from the partial hibernation of its intertube axial motion, the relative rotation slows down and is depleted of energies. Efforts have also been made to study DWNT oscillators of other chirality combinations, and similar results are found. This points to the universality of the inter-mode energy transfers amongst low-frequency mechanical modes of the DWNT oscillators, intertube axial motion included, regardless of chiralities of their composing SWNTs.

Our calculations on (5,0)/(14,0), (7,7)/(12,12) and other DWNT oscillators reveal that a DWNT oscillator with thousands of degrees of freedom can be reduced to a simple system with a few most relevant degrees of freedom in the presence of a thermal bath. Those few degrees of freedom correspond to several important low-frequency mechanical modes such as intertube axial oscillation, intertube rotation and bending-waving modes while the thermal bath is made of other higher-frequency vibrations of the nanotube. When the energy leakage from the reduced system to the bath is slow enough, the reversible energy exchange between the oscillator and the bending-rotational modes takes place thanks to the reduction of the effective phase space. A sketch of the reduced phase diagram is shown in Fig. 4 depicting energy transfers amongst three quasi-stable regimes. Such a picture of non-ergodic behavior is believed to underlie the observed reversible energy exchanges in the DWNT system. The three quasi-stable regimes are linked to form an enlarged non-ergodic zone that allows energy exchanges amongst them but otherwise prohibits a rapid thermal equilibration to take place. An interesting feature of this enlarged non-ergodic zone is that energy exchanges amongst quasi-stable modes are statistical in nature, unlike those in an isolated mechanical system with a few degrees of freedom in which energy exchanges amongst its various modes are deterministic in nature, i.e., predetermined by the initial conditions. The reduced system depicted in Fig. 4 can thus be a stable, yet flexible one with ample complexities.

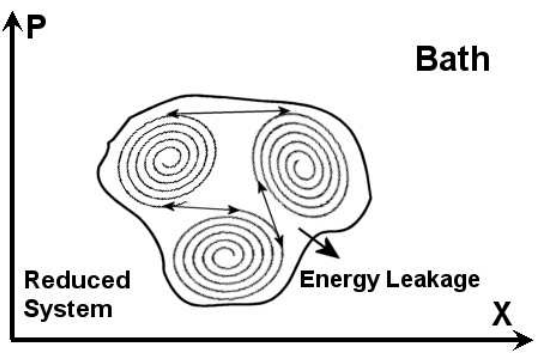

FIG. 4: The reduced phase diagram of a complex system. A: oscillation; B: bending-waving; and C: rotation.

In the literature dynamical behavior of identical particles interacting via the Lennard-Jones potential has been examined [14] aiming to investigate the characteristics of the classical phase space of coupled oscillators. DWNTs are in fact an extension of the two-dimensional Lennard-Jones model to a practical three-dimensional construct. Revelations and feedbacks from studies of statistical mechanics, in turn, can help design nanoscale mechanical devices. As conjectured by Sokoloff [15], a transition from frictional behavior to nearly frictionless sliding would occur as the size of the system decreases beyond a critical value. Sokoloff's conjecture is in fact an extreme case of the FPU persistent recurrence to the initial condition. It predicts the failure of the statistical equipartition at a certain small system size. What we have witnessed in this work is an intermediate case between the macroscopic system in which ergodicity holds and a nanoscale system in which Sokoloff's conjecture can be realized. This opens up the possibility of nearly frictionless and super-efficient nanoscale molecular oscillators in which practically no dissipation of the oscillator energies occur for a prolonged period of time.

Nonlinear modes have bands of amplitude-dependent frequencies, and when applied a driving force of a frequency which lies within this range of resonant frequencies, the nonlinear modes will not continue to absorb energy without limit. Rather, it will absorb energy up to a point after which the energy will flow back and forth between the driving force and the mode [16]. This tendency of anharmonic modes is demonstrated by our gigahertz nano-oscillator. The back-and-forth flows of energies in the tidal exchange can be attributed primarily to the driving force, which is powered by the intertube axial oscillation, and noncoaxial mechanical modes of a frequency circa 0.5 THz together with intertube rotational modes. Here the driving force has a frequency which is approximately the intertube axial speed divided by the size of the carbon hexagon. This frequency falls in the range of those of several important low-frequency mechanical modes of the DWNT system (including aforementioned modes). What is counterintuitive is that this reversible energy exchange phenomenon takes place between the orderly intertube axial oscillatory motion and mechanical modes in DWNT oscillators. Intertube axial oscillation of the DWNT oscillator is supposedly dissipated by frictional forces into disorderly phonons, and at the crossroad of the energy dissipation process lies this reversible energy exchange between the excess intertube van der Waals energy and low-frequency DWNT mechanical modes. We have shown in this Letter that the low-frequency vibrational-rotational modes can feed energy back into the intertube axial oscillatory motion resulting in a negative frictional force for the DWNT oscillator.

*Acknowledgments.* Support from the Hong Kong Research Grant Council (HKU 7012/04P) and the Committee for Research and Conference Grants (CRCG) of the University of Hong Kong is gratefully acknowledged. The authors thank L.H. Wong for assistance.

*Email: ghc@everest.hku.hk
†Email: zhengqs@tsinghua.edu.cn